\documentclass[conference]{IEEEtran}
\IEEEoverridecommandlockouts
\usepackage[style=ieee,url=false,doi=false,isbn=false]{biblatex}
\usepackage{amsmath,amssymb,amsfonts}
\usepackage{graphicx}
\usepackage{textcomp}
\usepackage{xcolor}
\usepackage{braket}
\usepackage{physics}
\usepackage{subfigure}
\usepackage{algorithm}
\usepackage{algpseudocode}
\usepackage{bm}
\def\BibTeX{{\rm B\kern-.05em{\sc i\kern-.025em b}\kern-.08em
    T\kern-.1667em\lower.7ex\hbox{E}\kern-.125emX}}

\begin{document}

\title{
  Transferring linearly fixed QAOA angles: performance and real device results
  \thanks{
    This work was performed for Council for Science, Technology and Innovation (CSTI), Cross-ministerial Strategic Innovation Promotion Program (SIP), ``Promoting the application of advanced quantum technology platforms to social issues'' (Funding agency: QST).
  }
}

\makeatletter
\newcommand{\linebreakand}{%
\end{@IEEEauthorhalign}
\hfill\mbox{}\par
\mbox{}\hfill\begin{@IEEEauthorhalign}
}
\makeatother

\author{
  \IEEEauthorblockN{
    Ryo Sakai,
    Hiromichi Matsuyama,
    Wai-Hong Tam,
    and Yu Yamashiro
  }
  \IEEEauthorblockA{
    \textit{Jij Inc.}, 3-3-6 Shibaura, Minato-ku, Tokyo, 108-0023, Japan
  }
  \IEEEauthorblockA{
    r.sakai@j-ij.com,
    h.matsuyama@j-ij.com,
    w.tam@j-ij.com,
    and y.yamashiro@j-ij.com
  }
}

\maketitle

\begin{abstract}
  Quantum Approximate Optimization Algorithm (QAOA) enables solving combinatorial optimization problems on quantum computers by optimizing variational parameters for quantum circuits.
  We investigate a simplified approach that combines linear parameterization with parameter transferring,
  reducing the parameter space to just 4 dimensions regardless of the number of QAOA layers p,
  while traditional QAOA requires optimizing 2p parameters that grow linearly with circuit depth.
  This simplification draws inspiration from quantum annealing schedules providing both theoretical grounding and practical advantages.
  We compare this combined approach with standard QAOA and other parameter setting strategies such as INTERP and FOURIER, which require computationally demanding incremental layer-by-layer optimization.
  Notably, previously known methods like INTERP and FOURIER yield parameters that can be well fitted by linear functions, which supports our linearization strategy.
  Our analysis reveals that for the random Ising model, cost landscapes in this reduced parameter space demonstrate consistent structural patterns across different problem instances.
  Our experiments extend from classical simulation to actual quantum hardware implementation on IBM's Eagle processor, demonstrating the approach's viability on current NISQ devices.
  Furthermore, the numerical results indicate that parameter transferability primarily depends on the energy scale of problem instances, with normalization techniques improving transfer quality.
  Most of our numerical experiments are conducted on the random Ising model,
  while problem-dependence is also investigated across other models.
  A key advantage of parameter transferring is the complete elimination of instance-specific classical optimization overhead,
  as pre-trained parameters can be directly applied to other problem instances,
  reducing classical optimization costs by orders of magnitude for deeper circuits.
  This approach offers a practical pathway for implementing QAOA on near-term quantum devices with reduced classical processing requirements.
\end{abstract}


\section{Introduction}
\label{sec:intro}

The quest for quantum utility in practical applications of variational quantum algorithms faces a significant obstacle:
as quantum circuits grow in complexity, the classical optimization overhead often becomes prohibitively expensive.
This challenge is particularly evident in the Quantum Approximate Optimization Algorithm (QAOA)~\cite{Farhi:2014ych,Blekos:2023nil,Abbas:2023agz}, where the parameter space grows linearly with circuit depth, creating a substantial classical optimization burden that limits practical deployment.

This paper presents a radical simplification to the QAOA paradigm through two complementary innovations.
First, we demonstrate that constraining QAOA parameters to strict linear functions with respect to circuit layer index dramatically reduces the parameter space from $2p$ dimensions to just $4$ dimensions, independent of the number of layers $p$.
Second, we establish that these linearized parameters exhibit remarkable transferability between problem instances, enabling zero-cost application to new problems without instance-specific optimization.

Parameter linearization draws inspiration from adiabatic quantum computing schedules~\cite{Farhi:2000ikn,Kadowaki:1998hua,Sack2021quantumannealing}, where linear parameter trajectories often yield near-optimal performance.
Our approach formalizes this intuition by imposing linearity as a strict constraint rather than an initialization strategy.
The resulting parameter structure creates a dramatically simplified optimization landscape with consistent structural properties across diverse problem instances.

Our work provides a comprehensive evaluation of linearly constrained QAOA parameters across various system sizes, problem types, and circuit depths.
We present empirical evidence that leading parameter setting strategies like INTERP and FOURIER~\cite{Zhou:2018fwi} naturally converge toward nearly-linear parameter patterns, lending credibility to our explicit linearization approach.
Through extensive numerical experiments, we demonstrate that parameters optimized for one problem instance can be directly transferred to other instances with minimal performance degradation.
We validate our approach on IBM's Eagle quantum processor, showing it maintains effectiveness despite realistic noise, and provide detailed analysis of parameter transferability with respect to energy scales~\cite{Shaydulin:2022ehb,Sureshbabu:2023tqu} and problem characteristics, developing a technique to enhance transfer quality.

Our findings reveal that linearized QAOA parameters consistently achieve approximation ratios comparable to fully optimized parameters while eliminating instance-specific optimization overhead.
For practitioners, this means QAOA can be applied to new problem instances with zero additional classical optimization cost---a transformative capability for quantum algorithm deployment on near-term devices.

This top-down approach to QAOA parameter optimization represents a departure from the conventional focus on maximizing approximation ratios for individual instances.
Instead, we prioritize practical deployment concerns: reducing classical overhead, enabling scalability to deeper circuits, and facilitating application to large problem instances without prohibitive optimization costs.

The remainder of this paper proceeds as follows: Sec.~\ref{sec:method_model} outlines the theoretical foundation of our approach.
Section~\ref{sec:result} presents experimental results encompassing method comparisons, landscape analyses, transferability studies, and problem dependency analyses.
Finally, Sec.~\ref{sec:summary} discusses implications and future research directions.

\section{Method and model}
\label{sec:method_model}

The central thesis of our approach is that QAOA parameters can be effectively constrained to linear functions of layer index, drastically reducing the parameter space while maintaining algorithm performance.
Before presenting our linearization strategy in detail, we first review the fundamental structure of QAOA and describe the optimization problems we target.

\subsection{Quantum Approximate Optimization Algorithm}
\label{sec:qaoa}

QAOA addresses combinatorial optimization by translating problems into finding the ground state of a cost Hamiltonian~\cite{Farhi:2014ych}.
Given a binary optimization problem with cost function $C(z)$, where $z \in \{0,1\}^{n}$, QAOA constructs a corresponding Hamiltonian whose ground state encodes the optimal solution.

The algorithm alternates between two operations: evolution under the problem Hamiltonian $e^{-i \gamma_{l} C}$ and evolution under a mixer Hamiltonian $e^{-i \beta_{l} H_{\mathrm{mix}}}$, typically chosen as $H_{\mathrm{mix}} = \sum_{j} X_{j}$ to enable exploration of the solution space~\footnote{
  For more sophisticated, problem-dependent mixer choices, see \textit{e.g.} Quantum Alternating Operator Ansatz~\cite{Hadfield:2017yqz}.
}.
Starting from the equal superposition state $\ket{+}^{\otimes n}$, QAOA applies $p$ layers of these evolution operators and then creates the parameterized state
\begin{equation}
  \Ket{\bm{\gamma}, \bm{\beta}} = e^{-i\beta_{p-1} H_{\mathrm{mix}}} e^{-i\gamma_{p-1} C} \cdots e^{-i\beta_{0} H_{\mathrm{mix}}} e^{-i\gamma_{0} C} \Ket{+}^{\otimes n}
\end{equation}

The algorithm's objective is to minimize the expectation value $\Braket{\bm{\gamma}, \bm{\beta} | C | \bm{\gamma}, \bm{\beta}}$ by optimizing the $2p$ parameters $\{\gamma_{l}, \beta_{l}\}_{l=0}^{p-1}$.
This hybrid quantum-classical approach measures the cost function on quantum hardware while a classical optimizer adjusts the parameters iteratively.

Traditional QAOA implementations treat each $\gamma_l$ and $\beta_l$ as independent, resulting in a $2p$-dimensional parameter space that grows linearly with circuit depth.
This expansion creates two significant challenges: the classical optimization becomes increasingly expensive, and the landscape develops more local minima that can trap optimization algorithms.

\subsection{Linear parameter simplification and transfer}
\label{sec:linxfer}

Our approach is based on a fundamental constraint:
QAOA parameters must follow strictly linear trajectories with respect to the layer index~\footnote{
  Apart from focusing on parameter concentration and transferability, the linear ramp QAOA was also studied in~\cite{2021arXiv210813056K,Kremenetski:2023dof,Shaydulin:2020jdg}.
}~\footnote{
  For more expressive but computationally demanding constrained parameters, see \textit{e.g.}~\cite{Wu:2023ddl}.
}.
Specifically, we parameterize as
\begin{align}
  \label{eq:lin_params}
  \begin{split}
    \gamma_{l} & = \gamma_{\mathrm{slope}} \frac{l}{p} + \gamma_{\mathrm{intcp.}}, \\
    \beta_{l} & = \beta_{\mathrm{slope}} \frac{l}{p} + \beta_{\mathrm{intcp.}},
  \end{split}
\end{align}

This constraint reduces the entire parameter space to just four dimensions ($\gamma_{\mathrm{slope}}$, $\gamma_{\mathrm{intcp.}}$, $\beta_{\mathrm{slope}}$, $\beta_{\mathrm{intcp.}}$), regardless of the number of layers $p$.
The slope parameters control how rapidly the angles change across layers, while the intercept parameters establish the starting values.

Several theoretical or observational considerations motivate this linearization.
First, in the $p \to \infty$ limit, QAOA converges to quantum adiabatic evolution, where linear parameter schedules are often effective~\cite{Sack2021quantumannealing}.
Second, when examining optimal parameters from QAOA, particularly those from sophisticated approaches like INTERP and FOURIER~\cite{Zhou:2018fwi}, we observe they frequently approximate linear functions naturally.
This feature is actually demonstrated in the numerical section in this paper.

Unlike previous works that use linear parameters as initial values for further optimization~\cite{Sack2021quantumannealing,Montanez-Barrera:2024pax}, we maintain this strict linear constraint throughout.
This imposes a strong structural prior on the parameter space, potentially limiting expressivity but dramatically simplifying optimization and enabling parameter transferability.

To identify optimal linear parameters for a reference instance, we employ Bayesian optimization via Optuna~\cite{10.1145/3292500.3330701}, exploring the four-dimensional parameter space efficiently.
Once optimized for one instance, the parameter set can be transferred to other instances.
This parameter setting strategy relies on the findings about parameter concentration reported in various studies~\cite{akshay2021parameter,Brandao:2018qoa,Farhi:2019xsx,Zhang:2024zbz}.
In a previous work~\cite{arxiv2405}, the authors have investigated some features of this parameterization, including visualizing typical structure in the energy landscape and transfer quality among small instances.

For later use in the numerical section, we define the linearized parameter transferring as in Alg.~\ref{alg:linxfer}.

\begin{algorithm}
  \caption{LINXFER: linearized parameter transferring}
  \begin{algorithmic}[1]
    \Require Source Hamiltonian $H$, number of layers $p$
    \Ensure Optimized linear QAOA parameters $\gamma$ and $\beta$

    \Function{Set}{$\gamma_{\mathrm{slope}}$, $\gamma_{\mathrm{intcp.}}$, $\beta_{\mathrm{slope}}$, $\beta_{\mathrm{intcp.}}$}
      \State $\gamma \gets [\:]$, $\beta \gets [\:]$ \Comment{Initialize empty lists}
      \For{$l = 0$ to $p-1$}
        \State $\gamma_l \gets \gamma_{\mathrm{slope}} \cdot (l/p) + \gamma_{\mathrm{intcp.}}$
        \State $\beta_l \gets \beta_{\mathrm{slope}} \cdot (l/p) + \beta_{\mathrm{intcp.}}$
        \State Add $\gamma_l$ to $\gamma$ and $\beta_l$ to $\beta$
      \EndFor
      \State \Return $\gamma, \beta$
    \EndFunction

    \Function{OptimizeLinear}{{}}
      \State $f(\gamma_{\mathrm{slope}}, \gamma_{\mathrm{intcp.}}, \beta_{\mathrm{slope}}, \beta_{\mathrm{intcp.}})$: $\expval{H}$ using $\gamma, \beta \gets$ \Call{Set}{$\gamma_{\mathrm{slope}}$, $\gamma_{\mathrm{intcp.}}$, $\beta_{\mathrm{slope}}$, $\beta_{\mathrm{intcp.}}$}
      \State \Return \Call{Bayesian}{$f$} \Comment{Optuna with $1024$ trials}
    \EndFunction

    \State $\gamma_{\mathrm{slope}}, \gamma_{\mathrm{intcp.}}, \beta_{\mathrm{slope}}, \beta_{\mathrm{intcp.}} \gets$ \Call{OptimizeLinear}{{}}
    \State $\gamma, \beta \gets$ \Call{Set}{$\gamma_{\mathrm{slope}}$, $\gamma_{\mathrm{intcp.}}$, $\beta_{\mathrm{slope}}$, $\beta_{\mathrm{intcp.}}$}
    \State \Return $\gamma, \beta$
  \end{algorithmic}
  \label{alg:linxfer}
\end{algorithm}

\subsection{Problem formulations}
\label{sec:models}

We focus primarily on a family of spin glass models, which provide a versatile framework for representing combinatorial optimization problems~\cite{Lucas:2013ahy}.
For a graph $G=(V,E)$ with vertices $V$ and edges $E$, the Ising~\cite{Lenz1920,Ising1925} Hamiltonian is
\begin{align}
  \label{eq:isinghamil}
  H = \sum_{(i,j) \in E} J_{ij} s_{i} s_{j},
\end{align}
where $s_i \in \{-1,+1\}$ represents a spin variable at vertex $i$, and $J_{ij}$ denotes the coupling strength between spins $i$ and $j$.
The optimization objective is to find the spin configuration that minimizes this energy function.

We investigate three primary problem classes:
\begin{enumerate}
\item \textbf{Random Ising model}: Each coupling $J_{ij}$ is randomly set to $+1$ or $-1$ with equal probability, creating a challenging optimization landscape with frustrated interactions.
\item \textbf{Max-cut problem}: A special case where all couplings are positive, equivalent to finding the maximum cut in a graph:
  \begin{align}
    \label{eq:maxcut}
    H_{\mathrm{maxcut}} = - \frac{1}{2} \sum_{(i,j) \in E} \left( 1 - s_{i} s_{j} \right),
  \end{align}
  with converted to a minimization problem and with constant factor and terms different from Eq.~\eqref{eq:isinghamil}.
\item \textbf{Sherrington--Kirkpatrick (SK) model}~\cite{Sherrington:1975zz}: Fully-connected graphs with normally distributed couplings $J_{ij}$, representing a canonical spin glass model.
\end{enumerate}

For all models, we characterize instances by the number of qubits $n_{\mathrm{qubits}} = |V|$ and the edge density $d_{\mathrm{edges}} = |E|/{}_{|V|}\mathrm{C}_{2}$, which represents the fraction of all possible edges present in the graph.

In our experimental framework, each spin variable maps directly to a qubit, allowing straightforward implementation on quantum hardware.
Our investigations span systems from small instances amenable to classical simulation up to larger systems approaching the capacity limits of current quantum processors.

\section{Result}
\label{sec:result}

We now present comprehensive empirical results validating the linearized parameter transferring.

\subsection{Comparative performance analysis}
\label{sec:comparison}

A fundamental question is whether constraining QAOA parameters to linear functions sacrifices too much algorithmic expressivity,
so to address the issue we now compare LINXFER (Alg.~\ref{alg:linxfer}) with established QAOA parameter setting strategies, INTERP and FOURIER~\cite{Zhou:2018fwi,Lee:2021hxh}, which can be seen as a family of layer-by-layer optimization algorithms~\cite{Lee:2023mhp,math11092176,Apte:2025niw}.
INTERP uses layer-by-layer optimization with interpolation between pre-optimized layers, while FOURIER parameterizes angles using Fourier series coefficients.
Both methods have shown strong performance but require significant computational overhead that increases with circuit depth.
Since it will be useful for keeping reproducibility and since they have some variants and options for \textit{e.g.} what classical optimization method is used, pseudocodes of both algorithms are shown in Appendix~\ref{sec:interp_fourier}.
Throughout this paper, we limit the number of Fourier terms used in FOURIER to $2$.
With this choice, tunable degrees of freedom in FOURIER turn to $4$ dimensional, just as in LINXFER.
Of course the more the number of Fourier terms is, the more the numerical accuracy should be,
but we adopt this choice just for the comparison purpose against LINXFER.
Results in this subsection are taken from a noiseless simulator Qulacs~\cite{Suzuki2021qulacsfast}.

For comparisons in this subsection, we use a parameter set that is optimized for a reference random Ising instance with $n_{\mathrm{qubits}}=16$ and $d_{\mathrm{edges}}=0.6$:
\begin{align}
  \label{eq:optparams_nqubits16_pedge0.6_nlayers8}
  \begin{split}
    \gamma_{l} & = - 0.376 \frac{l}{p} - 0.165, \\
    \beta_{l} & = - 0.881 \frac{l}{p} + 0.913,
  \end{split}
\end{align}
where $p=8$.
This parameter set was drawn from Bayesian optimization with $1024$ trials that took $153$ seconds on a laptop.
The parameter set has shown a nice transferability in the authors' previous work~\cite{arxiv2405} up to a $24$ qubit system that can be simulated by common state vector simulators~\footnote{
  Note that claiming a specialness of this specific parameter set is not our motivation.
  We sorely aim to demonstrate that the similarity of patterns in the energy landscape can be utilized for parameter transferring.
  Indeed, the optimal slopes and intercepts should naturally depend on the number of layers $p$,
  and it depends on the energy scale of instance as shown in a later subsection~\ref{sec:landscape}.
}.

A key insight emerges when examining the parameters produced by the different methods.
Figure~\ref{fig:params_comparison} shows the optimized parameters from standard QAOA~\footnote{
  For standard QAOA, we adopt uniform initialization so that every $\gamma$ and $\beta$ takes $0.1$.
  Note that random initialization in some range would be another option.
}, INTERP, FOURIER, and LINXFER for a $16$-qubit random Ising instance with $8$ QAOA layers.

\begin{figure}[htbp]
  \centering
  \includegraphics[width=\hsize]{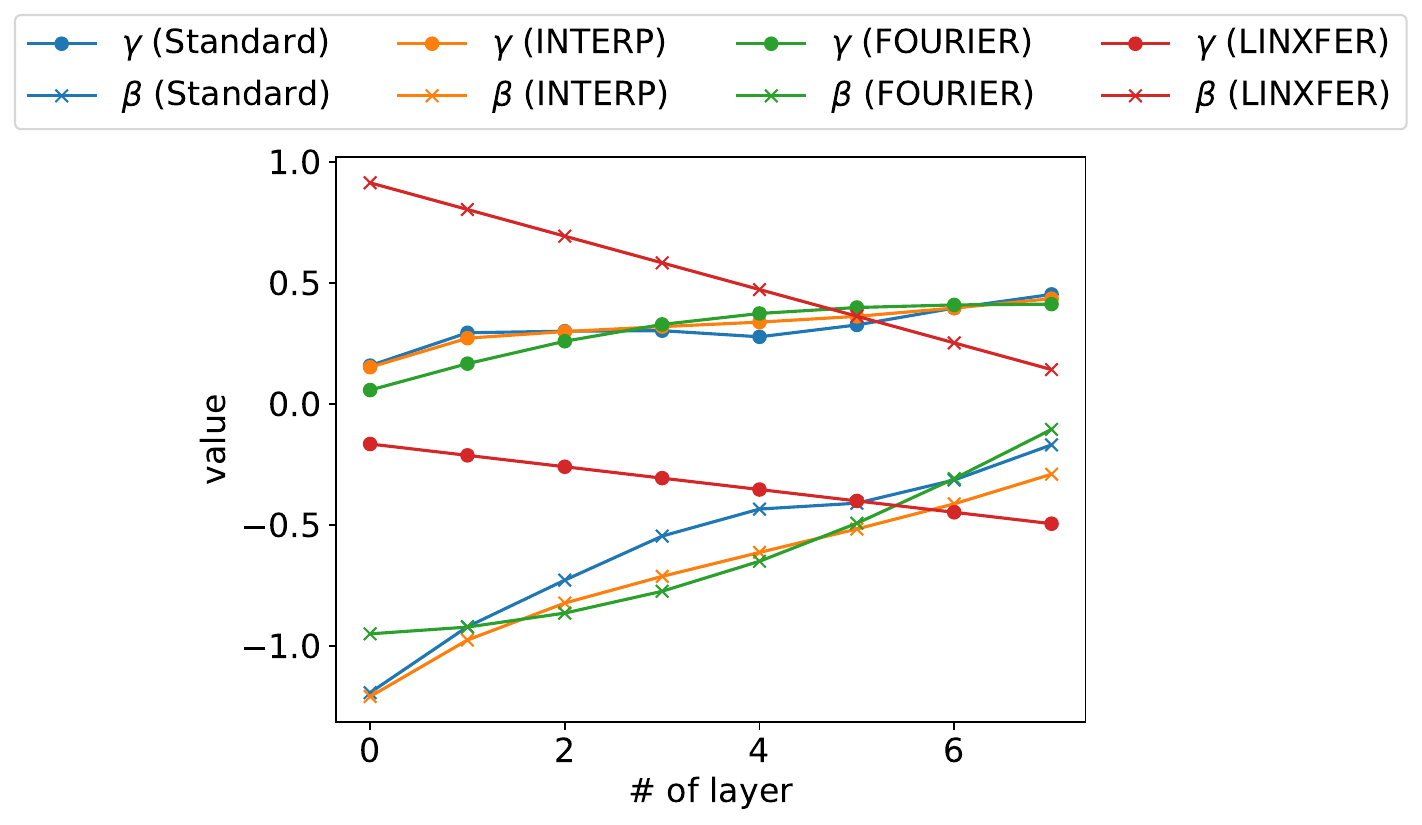}
  \caption{
    Comparison of optimized parameters from different methods for a $16$-qubit random Ising instance with $d_{\mathrm{edges}}=0.6$ and $p=8$.
    Parameters from standard QAOA, INTERP and FOURIER can be well fitted by linear functions, supporting the validity of our explicit linearization approach.
    We exclusively used COBYLA method for optimizing in INTERP, FOURIER, and standard QAOA with the number of function evaluations is limited to $1000$ for all methods.
    Note that the parameters labeled LINXFER is not specialized to this instance.
  }
  \label{fig:params_comparison}
\end{figure}

Remarkably, parameters from INTERP, FOURIER, and even the standard QAOA change across layers smoothly,
so that a fitting by linear functions would be legitimate, although we do not explicitly enforce linearity.
Moreover, these three behave consistently with each other.
These observations provide strong empirical justification for the linearization approach---what these sophisticated methods naturally discover is simply formalized, but with dramatically reducing computational overhead.

To compare performance, Fig.~\ref{fig:energy_comparison} shows energy distributions achieved by the four methods on the same instance.
While standard QAOA and INTERP achieve slightly better distribution concentration at the minimum energy,
LINXFER maintains competitive performance with significantly reduced computational overhead compared to the others.
Note that, this is a successful case for the standard QAOA.
As most practitioners know, QAOA often gets trapped in pseudo optima or $1000$ function evaluations are not enough at all;
this fact applies both INTERP and FOURIER too although the layer-by-layer optimization approaches often makes the situation better.

\begin{figure}[htbp]
  \centering
  \includegraphics[width=\hsize]{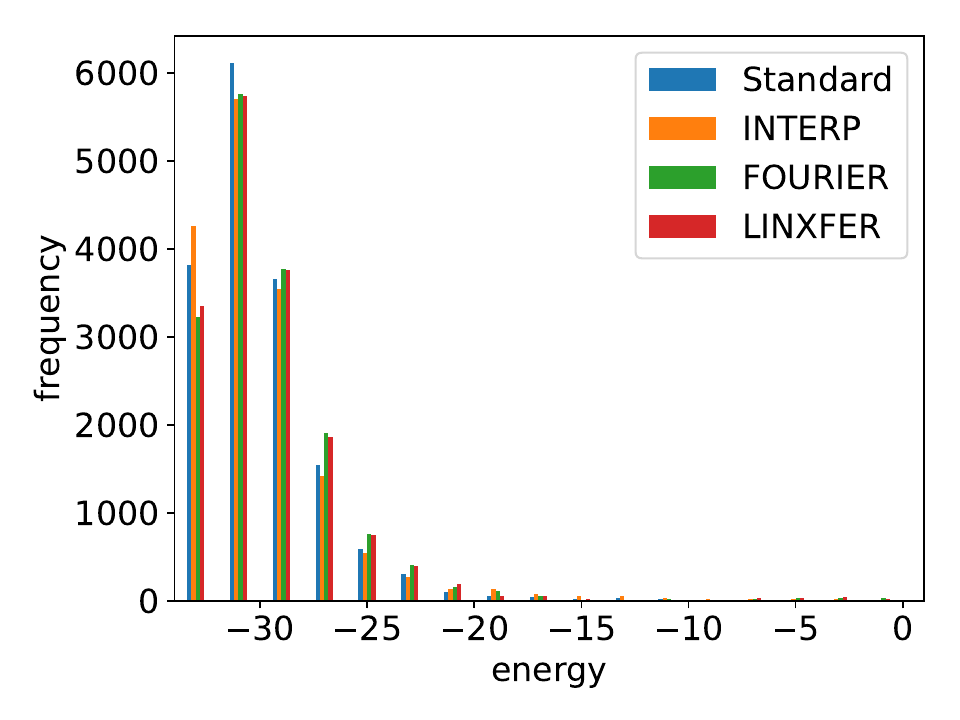}
  \caption{
    Energy distributions achieved by different parameter setting methods on a $16$-qubit random Ising instance with $d_{\mathrm{edges}}=0.6$ and $p=8$.
  }
  \label{fig:energy_comparison}
\end{figure}

The average of accuracy ($\expval{E}/E_{\mathrm{exact}}$) and of computational time are shown in Tables~\ref{tab:accuracy}--\ref{tab:comp_overhead} varying the number of layers $p$.
$8$ random instances where $(n_{\mathrm{qubits}}, d_{\mathrm{edges}}) = (16, 0.6)$ are used for taking the average for each $p$.

In terms of accuracy, the other three methods can be seen as slightly better than LINXFER,
but, when taking the statistical fluctuation into account, we can say all the methods are within error bars of each other for most $p$ cases.

Basically, the advantage of LINXFER in computational time becomes particularly pronounced for deeper circuits.
While other methods scale over linearly with circuit depth, our approach maintains zero cost regardless of the number of layers~\footnote{
  Sudden decrement of computational time at $p=16$ for Standard QAOA and FOURIER would be due to that the QAOA's expressive power gets much strong at $p=16$ for the difficulty of problem.
}.

These results demonstrate that our approach offers a compelling trade-off between solution quality and computational efficiency.
For applications where rapid deployment across multiple problem instances is prioritized over squeezing out the last percentage points of solution quality, LINXFER provides significant practical advantages.

\begin{table}[htbp]
  \caption{
    Average solution quality $\expval{E}/E_{\mathrm{exact}}$ for different methods and the number of layers.
  }
  \centering
  \begin{tabular}{lccccc}
    \hline
    Method & $p=2$ & $p=4$ & $p=6$ & $p=8$ & $p=16$ \\
    \hline
    Standard & 0.62(4) & 0.77(4) & 0.85(1) & 0.91(2) & 0.94(2) \\
    INTERP & 0.58(3) & 0.75(4) & 0.84(1) & 0.90(1) & 0.97(1) \\
    FOURIER & 0.62(4) & 0.74(4) & 0.83(1) & 0.89(1) & 0.96(2) \\
    LINXFER & 0.56(4) & 0.72(4) & 0.82(1) & 0.86(2) & 0.91(2) \\
    \hline
  \end{tabular}
  \label{tab:accuracy}
\end{table}

\begin{table}[htbp]
  \caption{
    Average parameter optimization time (in seconds) for different methods and the number of layers.
    Pre-training of Eq.~\eqref{eq:optparams_nqubits16_pedge0.6_nlayers8} took $153$ seconds.
    These calculations were carried out on a laptop with a fixed CPU allocation.
  }
  \centering
  \begin{tabular}{lccccc}
    \hline
    Method & $p=2$ & $p=4$ & $p=6$ & $p=8$ & $p=16$ \\
    \hline
    Standard & 15(6)s & 39(5)s & 117(44)s & 608(141)s & 189(45)s \\
    INTERP & 12(4)s & 43(10)s & 242(55)s & 1194(471)s & 2449(279)s \\
    FOURIER & 14(4)s & 10(2)s & 28(15)s & 200(68)s & 87(37)s \\
    LINXFER & 0s & 0s & 0s & 0s & 0s \\
    \hline
  \end{tabular}
  \label{tab:comp_overhead}
\end{table}

\subsection{Reduced landscape}
\label{sec:landscape}

In our past work~\cite{arxiv2405}, we have shown that despite the instances being generated with different random seeds, $n_{\mathrm{qubits}}$, and $d_{\mathrm{edges}}$, the energy landscapes exhibit striking structural similarities.
A typical example is shown for a random Ising instance in Fig.~\ref{fig:nqubits10_dedges0.4_nlayers3_nshots128_landscape}~(a), where $n_{\mathrm{qubits}}=10$, $d_{\mathrm{edges}}=0.4$, and $p=3$.
This pattern consistency suggests that optimal parameters for one instance should perform reasonably well for others---the foundation of our parameter transferability concept.

The landscape structure persistence extends beyond ideal simulation to quantum hardware implementation.
Figure~\ref{fig:nqubits10_dedges0.4_nlayers3_nshots128_landscape}~(b) shows cost landscapes measured on an IBM's Eagle quantum processor for the same $10$-qubit random Ising instance.
Remarkably, even in the presence of hardware noise, the fundamental landscape structure remains recognizable.
The typical structural pattern persists, demonstrating that noise does not drastically alter the topology of optimization landscape.
This resilience is crucial for practical applications,
as it suggests the parameter transfer approach can maintain effectiveness when deployed on actual quantum devices despite their imperfections.

\begin{figure}[htbp]
  \subfigure[Landscape on a noiseless simulator.]{
    \begin{minipage}{0.49\hsize}
      \includegraphics[width=\hsize]{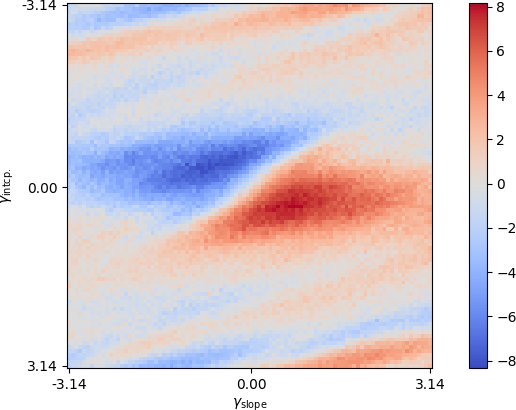}
    \end{minipage}
    \begin{minipage}{0.49\hsize}
      \includegraphics[width=\hsize]{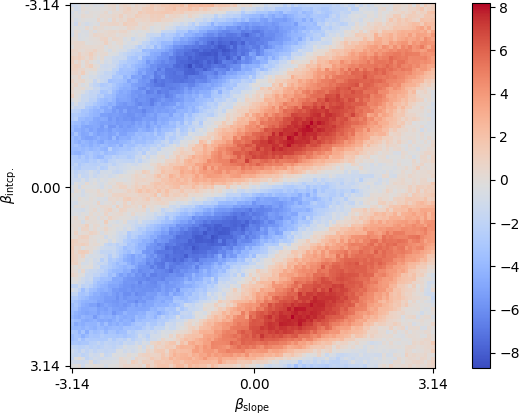}
    \end{minipage}
  }
  \subfigure[Landscape on ibm\_brisbane.]{
    \begin{minipage}{0.49\hsize}
      \includegraphics[width=\hsize]{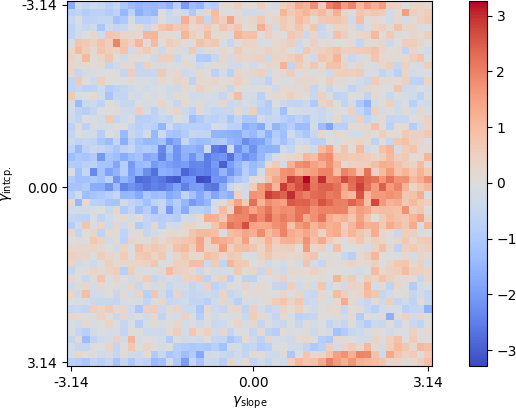}
    \end{minipage}
    \begin{minipage}{0.49\hsize}
      \includegraphics[width=\hsize]{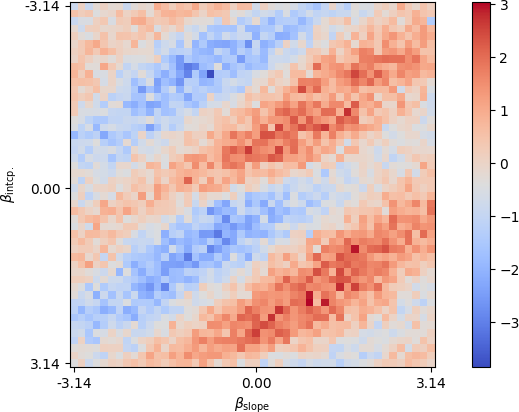}
    \end{minipage}
  }
  \caption{
    Cost landscapes measured on (a) noiseless simulator (Qulacs) and (b) IBM's Eagle quantum processor (ibm\_brisbane) for a $10$-qubit random Ising instance with $d_{\mathrm{edges}}=0.4$ and $p=3$.
    Despite hardware noise, the key structural features remain discernible, indicating the approach's resilience to realistic quantum noise.
    Q-CTRL's performance management software~\cite{Mundada:2022roq} provided through IBM Quantum services is applied.
  }
  \label{fig:nqubits10_dedges0.4_nlayers3_nshots128_landscape}
\end{figure}

An essential factor affecting landscapes is the energy scale of the problem instance~\cite{Shaydulin:2022ehb,Sureshbabu:2023tqu,arxiv2405}.
Here we directly observe that when the overall magnitude of Hamiltonian coefficients changes, the landscape undergoes a predictable transformation, particularly in the $\gamma$ parameter space.

Figure~\ref{fig:normalized_landscape} shows landscapes for a fixed instance with different normalization factors for each edge (or for overall Hamiltonian).
We normalize the edge weights by $|E_{\mathrm{SA}}|/X$ for $X \in \{8, 16, 32\}$, where $E_{\mathrm{SA}}$ is the (quasi) ground state energy that is obtained by simulated annealing using OpenJij~\cite{OpenJij}.

\begin{figure*}[htbp]
  \begin{minipage}{0.32\hsize}
    \includegraphics[width=\hsize]{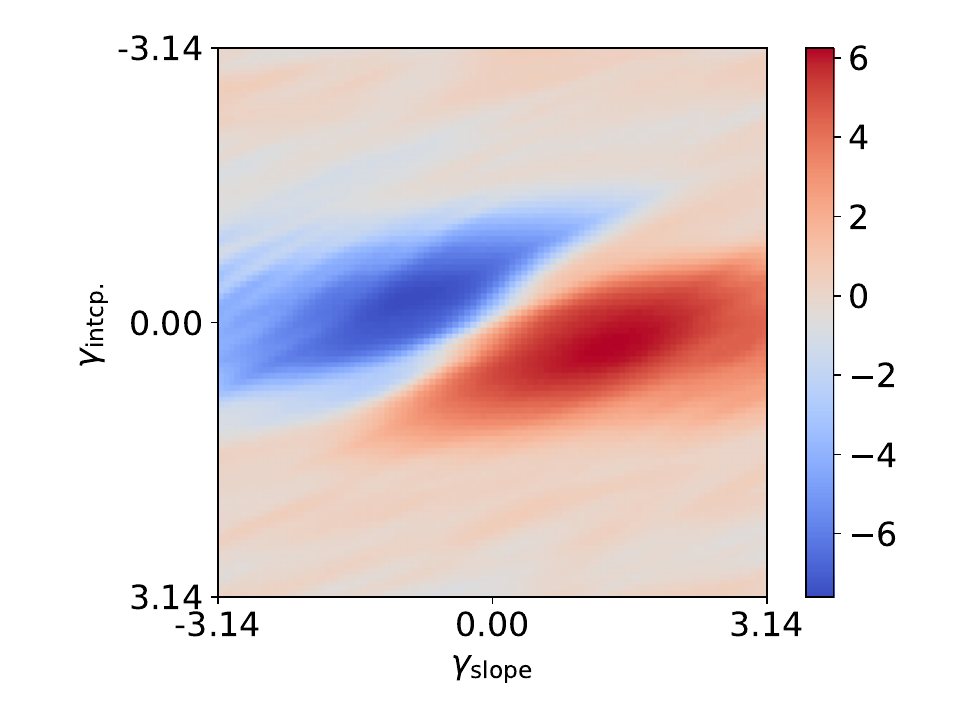}
  \end{minipage}
  \begin{minipage}{0.32\hsize}
    \includegraphics[width=\hsize]{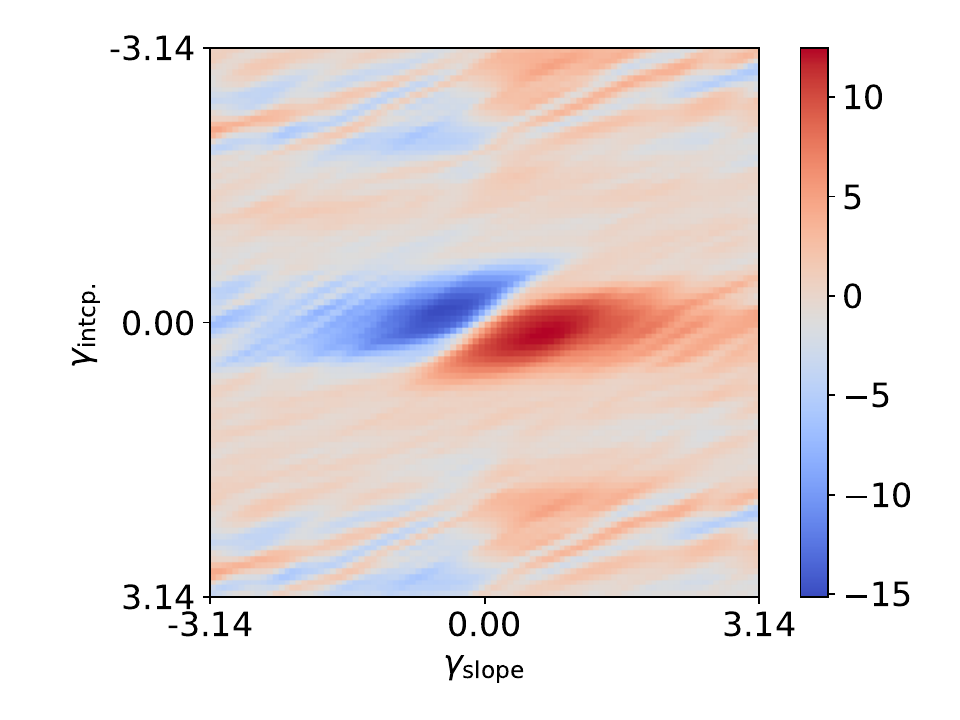}
  \end{minipage}
  \begin{minipage}{0.32\hsize}
    \includegraphics[width=\hsize]{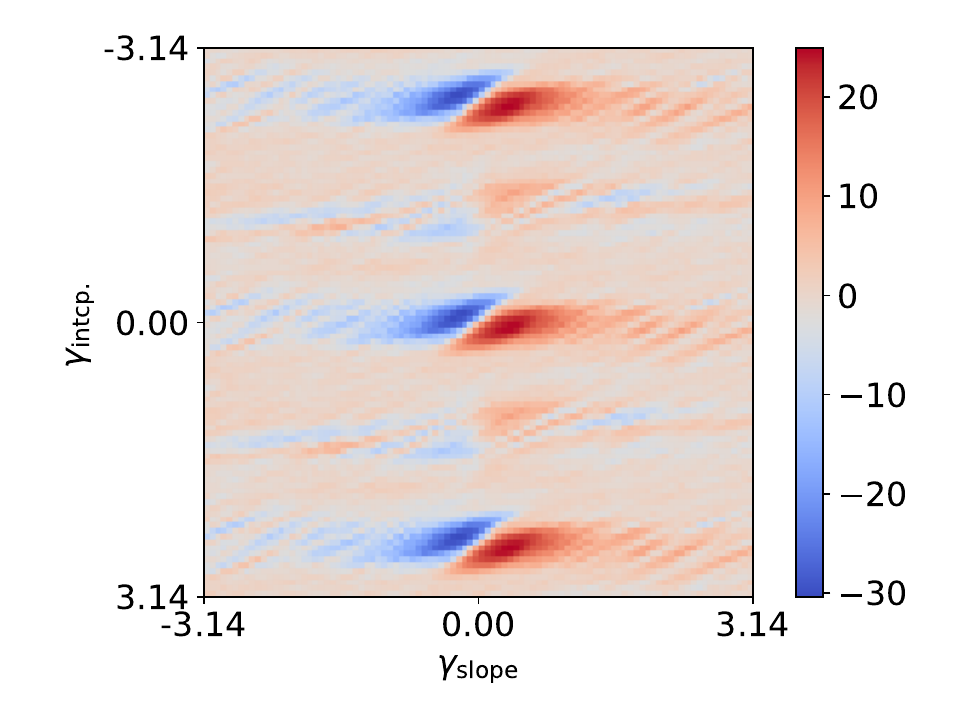}
  \end{minipage}
  \caption{
    Normalized energy landscapes in $(\gamma_{\mathrm{slope}}, \gamma_{\mathrm{intcp.}})$-space for a random Ising instance with $n_{\mathrm{qubits}}=9$ and $d_{\mathrm{edges}}=0.6$.
    The number of layers is set as $p=8$.
    The normalization factor for edges is taken to be $|E_{\mathrm{SA}}|/8$ (left), $|E_{\mathrm{SA}}|/16$ (center), and $|E_{\mathrm{SA}}|/32$ (right).
  }
  \label{fig:normalized_landscape}
\end{figure*}

The normalization for edges effectively adjusts the "zoom level" of the landscape, allowing us to transfer parameters between problem classes with inherently different energy scales.
This is natural since the normalization factor of Hamiltonian converts the periodicity in $\gamma$-plane;
multiplying the Hamiltonian by a factor corresponds to dividing the optimal $\gamma$ parameters by the same factor.

Figure~\ref{fig:best_gamma_scaling} shows the scaling of best $\gamma$ position with changing $X$ in the normalization factor $|E_{\mathrm{SA}}|/X$.
As $X$ increases, the best $\gamma$ position approaches to the origin in the $(\gamma_{\mathrm{slope}}, \gamma_{\mathrm{intcp.}})$-space.
A heuristic usage of this property is tuning $X$ to match the destination energy scale to the source one.
In practice, a Goemans--Williamson's approximate energy~\cite{10.1145/227683.227684} or the number of edges $n_{\mathrm{edges}}$ can be used instead of $|E_{\mathrm{SA}}|$.

\begin{figure}[htbp]
  \centering
  \includegraphics[width=\hsize]{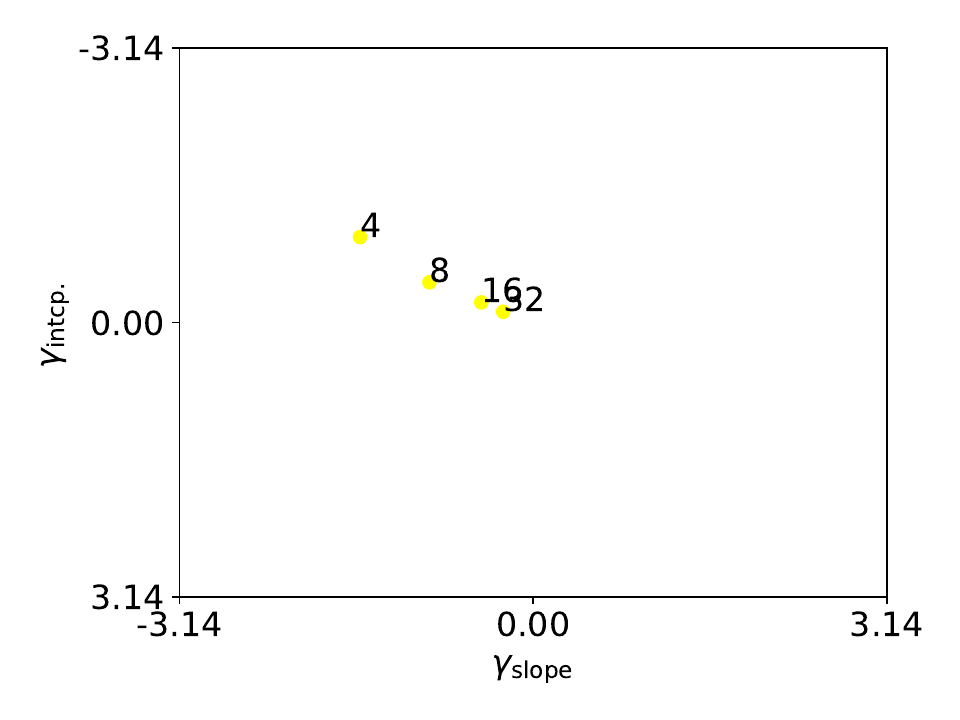}
  \caption{
    Scaling of best $\gamma$ with changing normalization factor $X$ in $|E_{\mathrm{SA}}|/X$.
  }
  \label{fig:best_gamma_scaling}
\end{figure}

\subsection{Problem size scaling of parameter transferability}
\label{sec:transfer_scaling}

A critical test of our approach is whether a given parameter set can be effectively transferred to large systems.
This capability is particularly valuable for practical applications,
as it enables addressing problem sizes that would be prohibitively expensive to optimize directly.

To demonstrate a scaling of transferability and to show the effectiveness of normalization suggested in the previous subsection,
we apply a toy parameter set
\begin{align}
  \label{eq:rough_guess}
  \begin{split}
    \gamma_{l} = -\frac{l}{p} - 1, \\
    \beta_{l} = -\frac{l}{p} + 1
  \end{split}
\end{align}
to large instances with normalizing the target Hamiltonian by $|E_{\mathrm{SA}}|/\sqrt{n_{\mathrm{edges}}}$.
Note that this experiment setting is quite toy-ish:
we set every slope and intercept to be unit~\footnote{
  Of course, we could use a Bayesian optimized parameter set for an actual instance like Eq.~\eqref{eq:optparams_nqubits16_pedge0.6_nlayers8} instead of Eq.~\eqref{eq:rough_guess};
  however, we use the latter to make the effect of normalization clearer.
}, and indeed, the normalization factor here is a super heuristic component.
One do not necessarily know $E_{\mathrm{SA}}$ and indeed $|E_{\mathrm{SA}}|/\sqrt{n_{\mathrm{edges}}}$ much fluctuates depending on instance,
so, in actual usages, one can use \textit{e.g.} the Goemans--Williamson's solution instead, as mentioned above.

Figure~\ref{fig:transfer_large} shows energy distributions for increasingly large random Ising instances up to $80$ qubits using transferred parameters.
Since the exact energies $E_{\mathrm{exact}}$ cannot be obtained straightforwardly in such large systems, we also show $E_{\mathrm{SA}}$ for comparison.
Note that state vector simulators cannot simulate such large systems, so that we make use of a matrix product states (MPS) simulator provided by IBM while we exclusively fix the bond dimension to be $32$ (The legitimacy of this choice is verified in Appendix~\ref{sec:bond_dim_dependence}).
Using real devices is of course a natural option,
but we figured out that the noise strength in the current devices masks the distinguishment between QAOA and random sampling in large systems~\footnote{
  Many other works (see \textit{e.g.} Ref.~\cite{Montanez-Barrera:2024tos}) apply local search algorithm to sampling results as error mitigation;
  \textit{i.e.} each bit is flipped as trial, and it is accepted if the energy value decreases.
  However, one has to be careful when applying such a mitigation approach since the local search is often able to solve the random Ising models and the max-cut problems by itself for problem sizes with $n_{\mathrm{qubits}} \sim 100$.
  Thus we have decided to use the MPS simulator here for putting our focus only on parameter transferability across different problem sizes rather than noise.
  Note that a robustness against hardware noise is fairly investigated in Sec.~\ref{sec:landscape}.
}.

The parameter set (Eq.~\eqref{eq:rough_guess}) for this experiment is quite roughly set with just assuming the linearity and never optimized for any actual problem instance.
Thus, the original sampling results without edge normalization (blue histograms in Fig.~\ref{fig:transfer_large}) behave like those from random sampling whose center is roughly located at $0$ energy.
With the edge normalization, however, the location of histogram is drastically improved, showing the power of proper normalization.

\begin{figure}[htbp]
  \subfigure[$n_{\mathrm{qubits}}=32$. $E_{\mathrm{SA}}=-103$]{
    \includegraphics[width=\hsize]{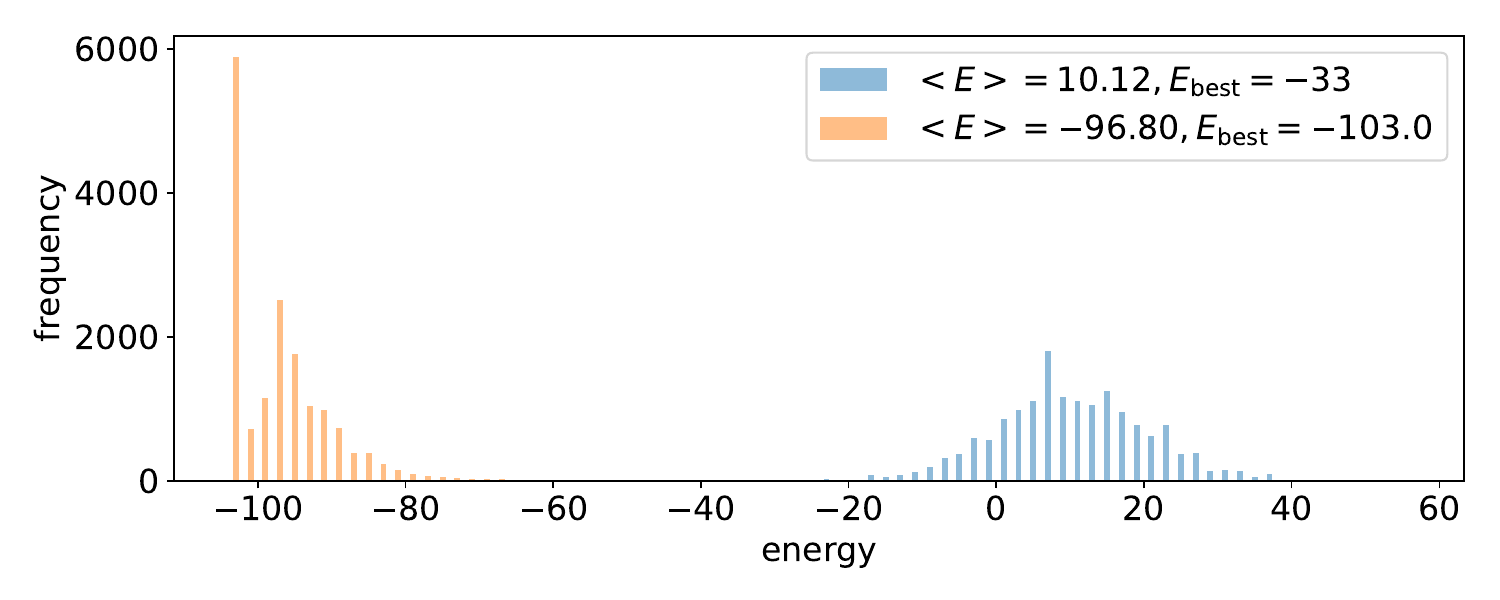}
  }
  \subfigure[$n_{\mathrm{qubits}}=48$. $E_{\mathrm{SA}}=-185$]{
    \includegraphics[width=\hsize]{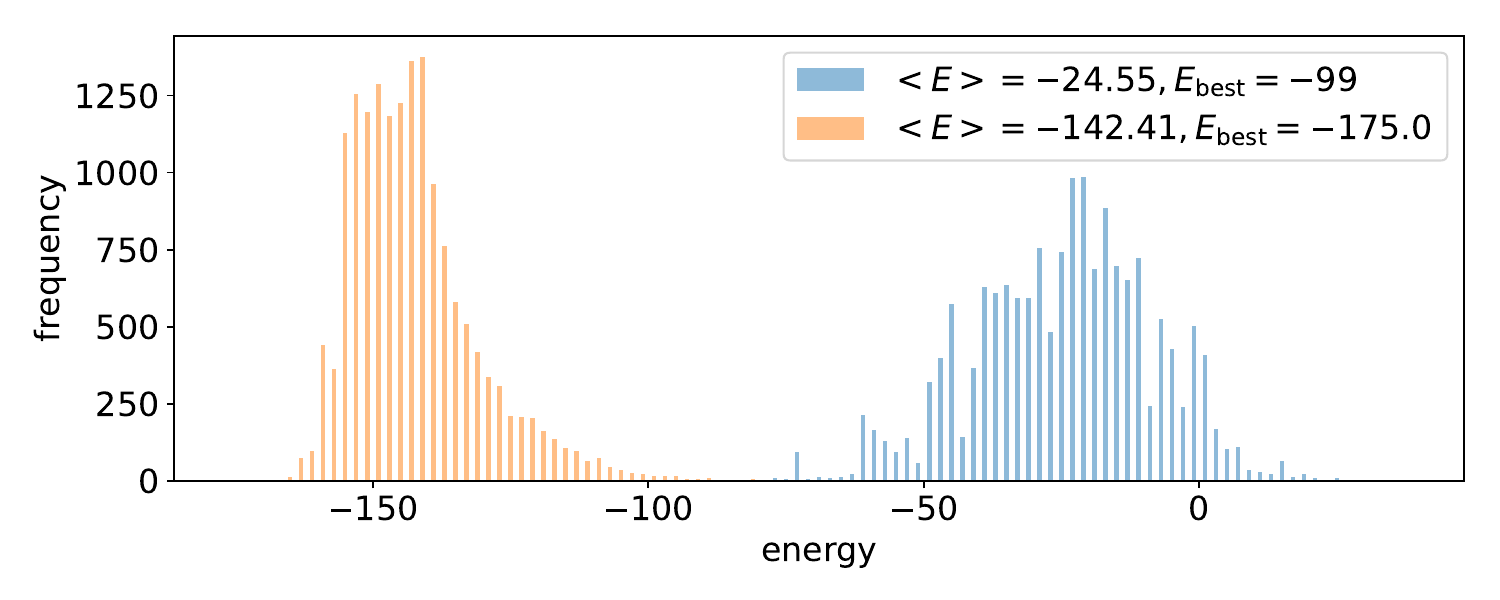}
  }
  \subfigure[$n_{\mathrm{qubits}}=64$. $E_{\mathrm{SA}}=-291$]{
    \includegraphics[width=\hsize]{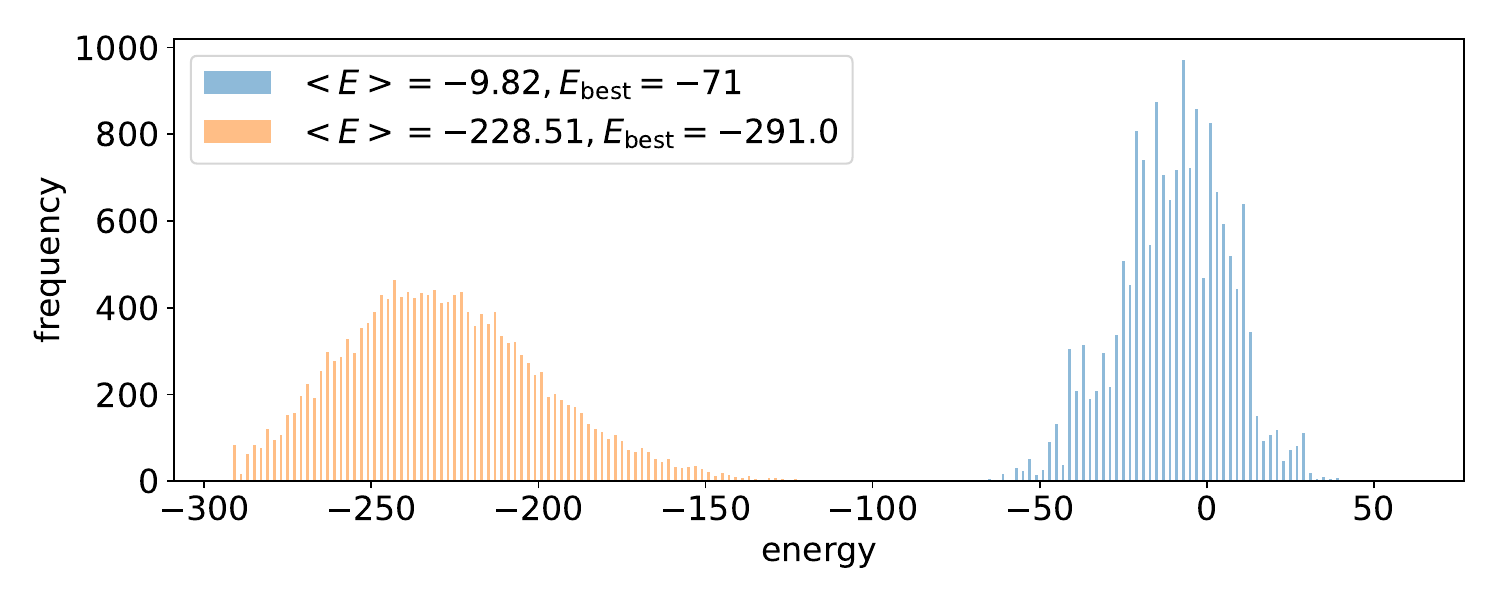}
  }
  \subfigure[$n_{\mathrm{qubits}}=80$. $E_{\mathrm{SA}}=-394$]{
    \includegraphics[width=\hsize]{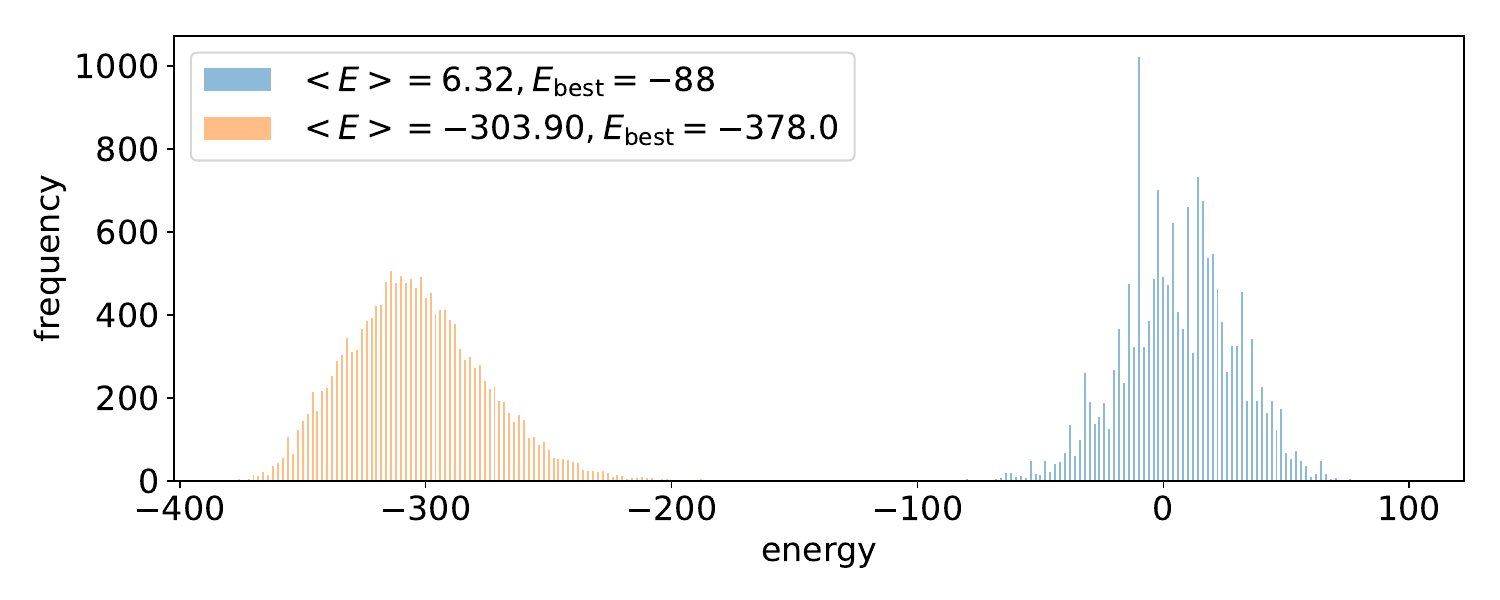}
  }
  \caption{
    Energy distribution histograms for transferred parameters on large random Ising instances with edge normalization (orange).
    For comparison the sampling results without edge normalization are also shown (blue).
    The edge weights are de-normalized when calculating the energy for display.
    $p=8$.
  }
  \label{fig:transfer_large}
\end{figure}

Figure~\ref{fig:approx_ratio_scaling} shows the approximation ratios as functions of system size with using the parameters (Eq.~\eqref{eq:rough_guess}) and the edge normalization by $|E_{\mathrm{SA}}|/\sqrt{n_{\mathrm{edges}}}$.
The transferred parameters maintain impressive performance, with approximation ratios $\expval{E}/E_{\mathrm{SA}}$ typically exceeding $0.8$ even for the $80$-qubit instances.
Also, $E_{\mathrm{best}}/E_{\mathrm{SA}}$ reaches a quality close to $1$ for all system size up to $n_{\mathrm{qubits}}=80$.
This demonstrates that our approach scales effectively to system sizes that would be impractical to optimize directly.

\begin{figure}[htbp]
  \centering
  \includegraphics[width=\hsize]{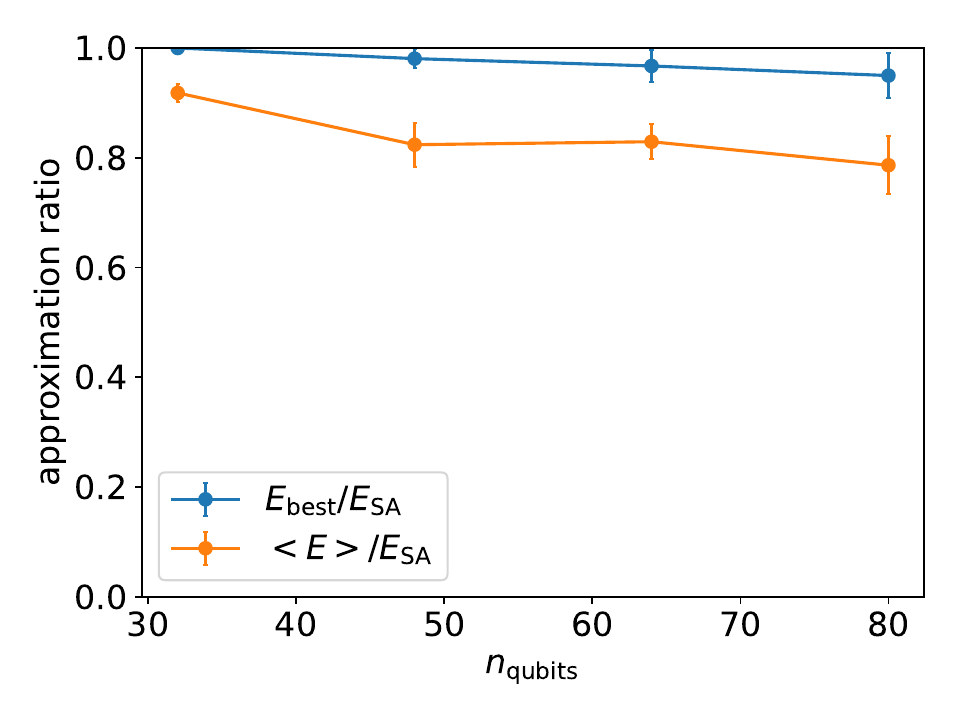}
  \caption{
    Scaling of parameter transferability with system size.
    The plot shows approximation ratios for random Ising instances with $d_{\mathrm{edges}}=0.6$ across different qubit counts using the MPS simulator.
    The bond dimension and the number of QAOA layers are set to $32$ and $8$, respectively.
    Each point represents an average over $8$ random instances.
    Performance remains consistently strong up to the largest tested systems with $n_{\mathrm{qubits}}=80$.
  }
  \label{fig:approx_ratio_scaling}
\end{figure}

It is worth noting that the MPS simulator introduces a different type of error than quantum hardware noise---specifically, errors arising from the truncated bond dimension in the tensor network representation.
These errors tend to increase with system size and entanglement, similar to how hardware errors accumulate in larger circuits, but with different underlying mechanisms.
The fact that parameter transferability remains robust even with these simulation-specific errors further supports the approach's resilience to various forms of noise and approximation.

These results demonstrate that our LINXFER approach effectively addresses one of the key challenges in practical quantum optimization:
enabling application to large problem instances without prohibitive classical optimization overhead~\footnote{
  For large systems where the QAOA optimization with state vector simulators is not an option, tensor network simulators are of course effective even though there is unavoidable truncation error.
  Indeed, in Ref.~\cite{OLeary:2025mdw}, an MPS surrogate model is used for training QAOA parameters on large systems, and transferring to real devices is also considered,
  although the linearity constraint of parameters is not imposed.
}.

The result in this section raises a new heuristic approach besides LINXFER:
one can regard the normalization factor for the Hamiltonian as a tunable parameter and can iteratively optimize it.
Such an approach does not need $E_{\mathrm{SA}}$ as prior knowledge, and the complexity of estimating the target energy scale will be replaced by the iteration overhead.

\subsection{Cross-problem analysis}
\label{sec:problem_transferability}

For practical quantum optimization deployment, parameter transferability should extend beyond similar instances of a single problem class.
In this subsection, we examine how the typical landscape structure is persistent across different problem formulations with particular attention to the SK models and the max-cut problems.

Figure~\ref{fig:sk_landscape} presents landscapes for instances of the SK model, which features normally distributed couplings on fully-connected graphs.
Interestingly, the SK model landscapes show a dependence on the width of the coupling distribution that parallels the energy scale effects discussed earlier in this paper.
Broader distributions effectively increase the energy scale, requiring corresponding adjustments to the $\gamma$ parameters.
This observation further reinforces the importance of energy scale normalization for cross-problem transferability.
These results highlight an important characteristic of our approach:
while the specific shape of optimal parameters depends on problem structure, the linearity constraint and energy scale normalization provide sufficient adaptability to maintain effectiveness across different problem classes.

\begin{figure*}[htbp]
  \subfigure[$(\gamma_{\mathrm{slope}}, \gamma_{\mathrm{intcp.}})$-space.]{
    \begin{minipage}{0.32\hsize}
      \includegraphics[width=\hsize]{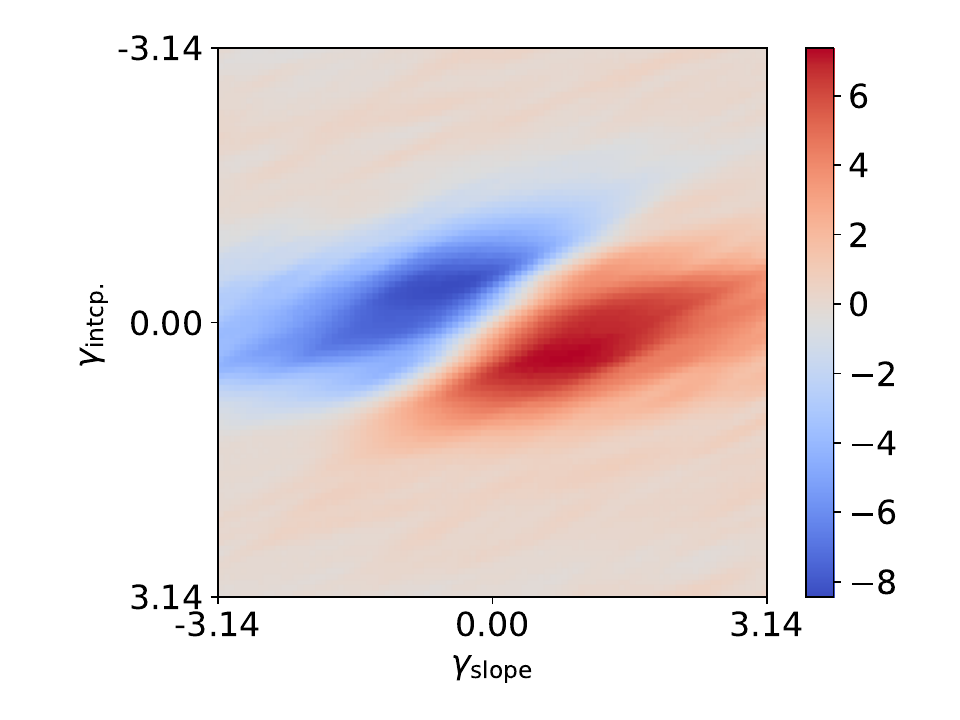}
    \end{minipage}
    \begin{minipage}{0.32\hsize}
      \includegraphics[width=\hsize]{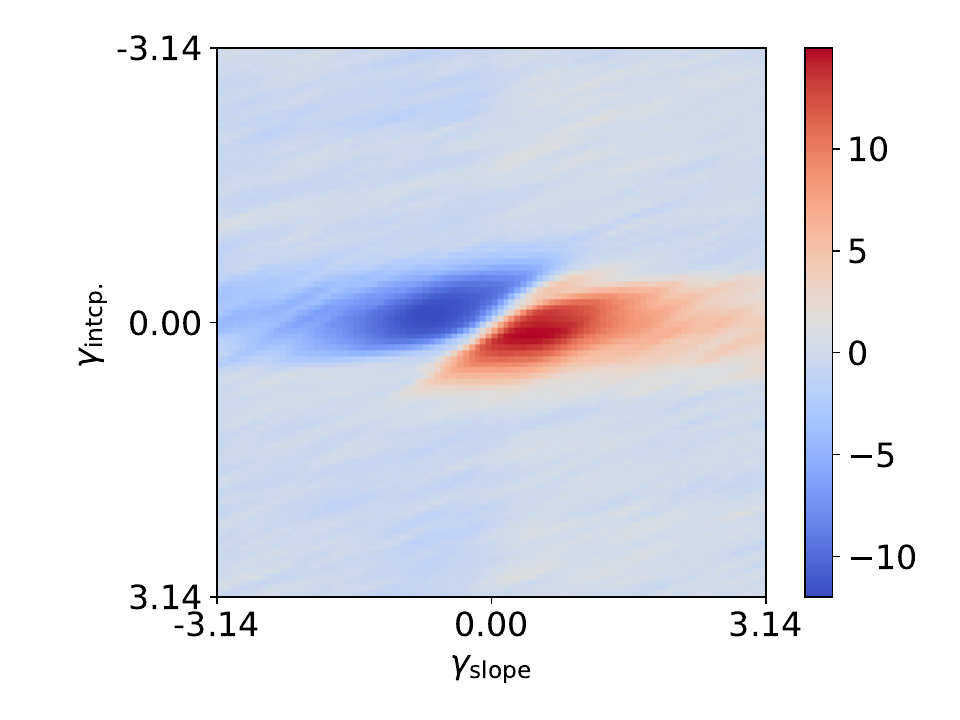}
    \end{minipage}
    \begin{minipage}{0.32\hsize}
      \includegraphics[width=\hsize]{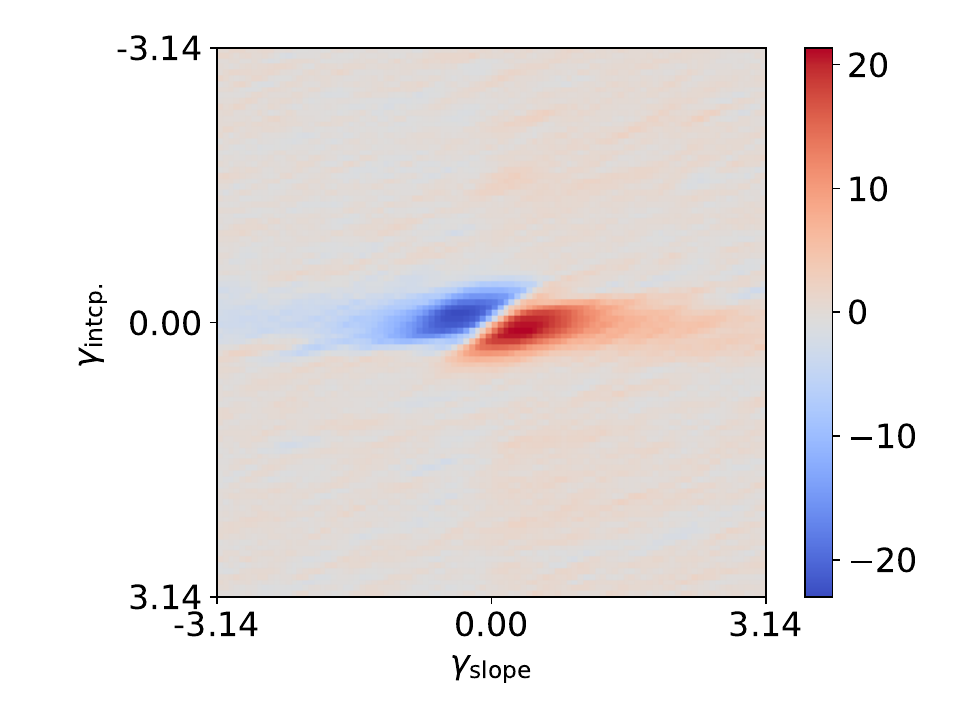}
    \end{minipage}
  }
  \subfigure[$(\beta_{\mathrm{slope}}, \beta_{\mathrm{intcp.}})$-space.]{
    \begin{minipage}{0.32\hsize}
      \includegraphics[width=\hsize]{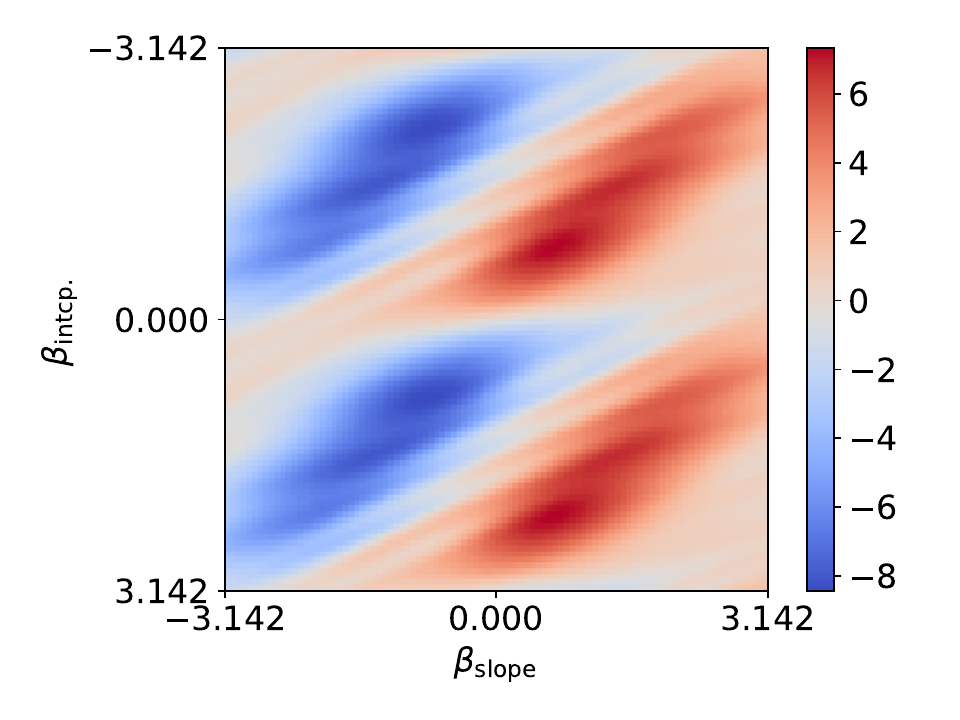}
    \end{minipage}
    \begin{minipage}{0.32\hsize}
      \includegraphics[width=\hsize]{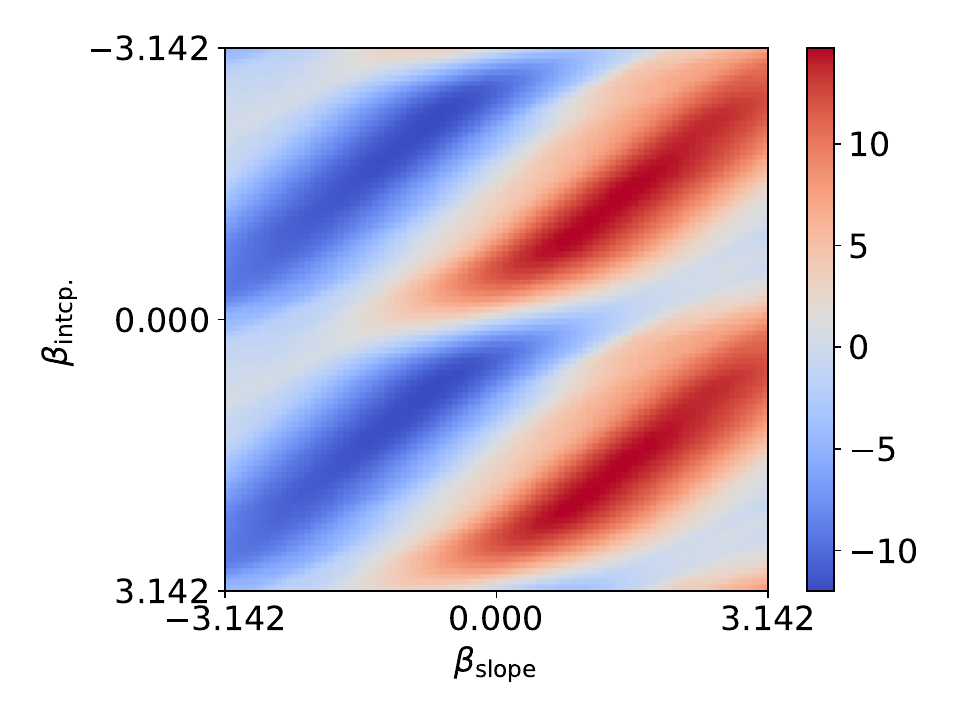}
    \end{minipage}
    \begin{minipage}{0.32\hsize}
      \includegraphics[width=\hsize]{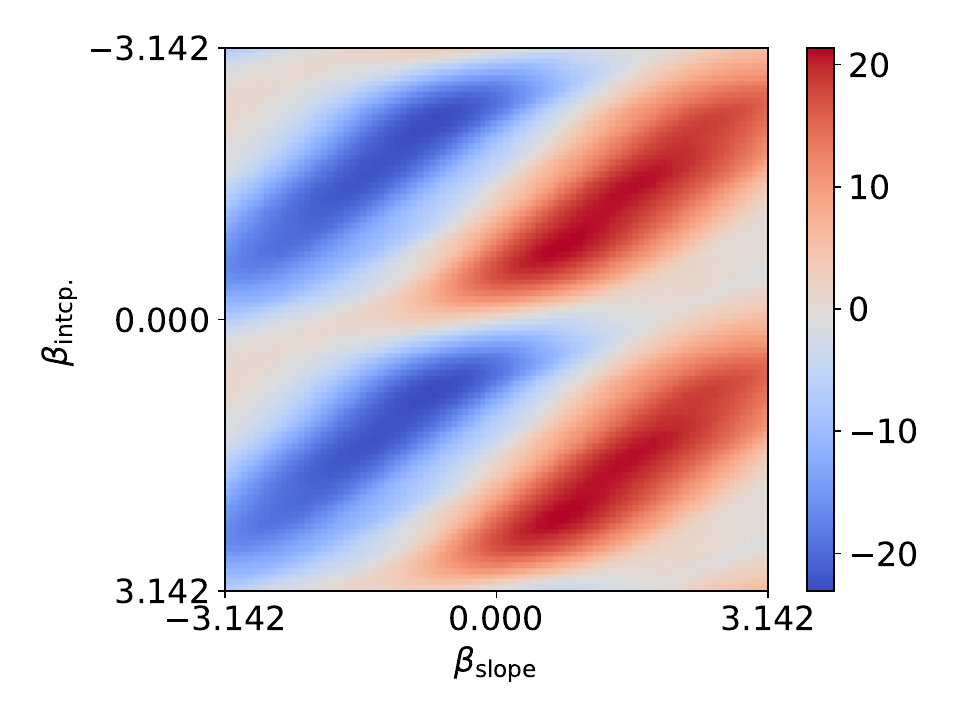}
    \end{minipage}
  }
  \caption{
    Cost landscapes for SK model with $9$ qubits: (a) $(\gamma_{\mathrm{slope}}, \gamma_{\mathrm{intcp.}})$- and (b) $(\beta_{\mathrm{slope}}, \beta_{\mathrm{intcp.}})$-space.
    The width of interaction distribution that affects the energy scale is varied as $J_{ij} \sim N(0, 4/n_{\mathrm{qubits}})$~(left), $N(0, 8/n_{\mathrm{qubits}})$~(center), and $N(0, 16/n_{\mathrm{qubits}})$~(right).
  }
  \label{fig:sk_landscape}
\end{figure*}

Note that these experimental results on the SK model are encouraging for future prospects towards real devices.
In QAOA, the evolution of problem Hamiltonian involves the $R_{ZZ}$ gates whose angles are the coupling coefficients $J_{ij}$,
so hardware errors will emerge as disturbance for each $J_{ij}$ independently, just like the probabilistic fluctuation of coupling coefficients in the SK model.
Thus, the stability of landscape pattern in the SK model indirectly supports a noise-robustness of our approach.
Moreover, our results in the previous subsection show that we can alter the zoom level of energy landscape by Hamiltonian normalization;
this fact implies that we can suppress noise effects (width of fluctuation, in the SK model) by choosing an appropriate scaling factor.
Of course there are many types of noise on real devices besides those on the $R_{ZZ}$ gates,
so in any case this intuition needs to be verified on actual devices.

We show a cost landscape for a randomly generated max-cut problem instance in Fig.~\ref{fig:maxcut_landscape}.
While the random Ising and the SK models have coupling constants that have plus and minus signs probabilistically, the coupling constant in the max-cut problem takes a uniform sign.
Even in such a fully-biased case, the figure shows the common landscape patterns as those in the random Ising and the SK models.

\begin{figure}[htbp]
  \begin{minipage}{0.49\hsize}
    \includegraphics[width=\hsize]{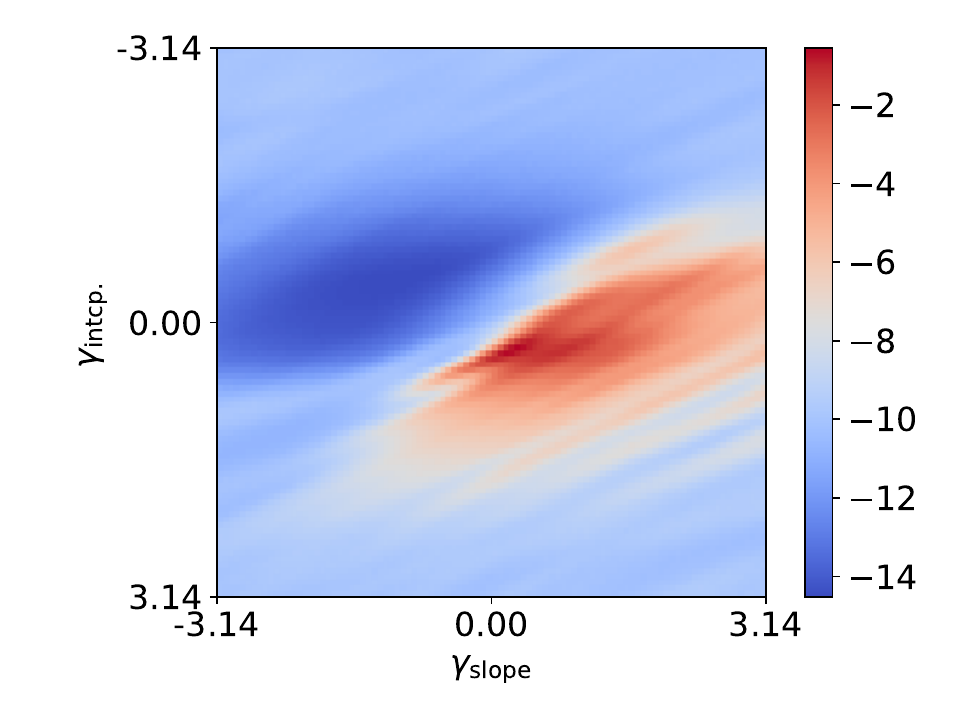}
  \end{minipage}
  \begin{minipage}{0.49\hsize}
    \includegraphics[width=\hsize]{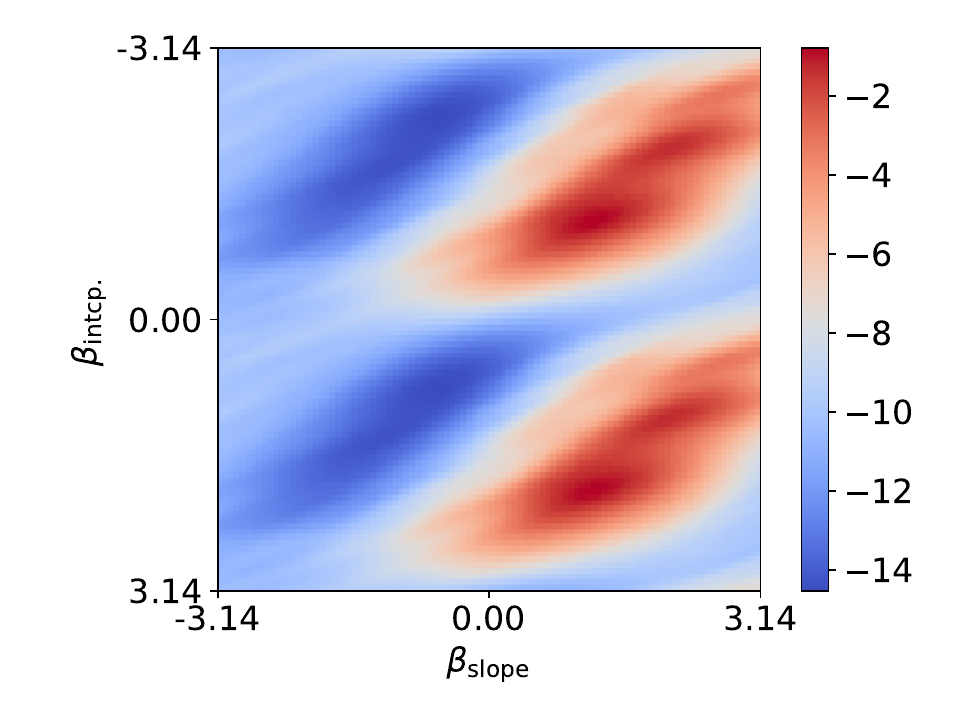}
  \end{minipage}
  \caption{
    Cost landscape for max-cut problem with $9$ qubits and $d_{\mathrm{edges}}=0.6$: $(\gamma_{\mathrm{slope}}, \gamma_{\mathrm{intcp.}})$- and $(\beta_{\mathrm{slope}}, \beta_{\mathrm{intcp.}})$-space.
  }
  \label{fig:maxcut_landscape}
\end{figure}

The cross-problem transferability discussed here significantly expands the practical utility of our approach.
By understanding how problem characteristics affect cost landscape and applying appropriate normalization, we can leverage parameters optimized for one problem class to address diverse optimization challenges across multiple problem formulations.

\section{Summary}
\label{sec:summary}

This work advances quantum optimization through a conceptual shift in how we approach QAOA parameter setting.
Rather than pursuing incremental improvements in approximation ratios through increasingly complex optimization procedures, we have demonstrated that dramatically simplifying the parameter structure enables practical advantages for near-term applications.

Our work establishes that constraining QAOA parameters to strict linear functions of layer index reduces the parameter space from $2p$ to just $4$ dimensions regardless of circuit depth, while maintaining performance competitive with fully optimized parameters.
This constraint sacrifices at most a few percentage points in approximation ratio while offering dramatic computational advantages.
Cost landscapes in the reduced four-dimensional parameter space exhibit remarkable structural similarity across different problem instances, despite varying system sizes and edge connectivities, providing the foundation for parameter transferability.

Experiment on IBM's Eagle quantum processor confirms that our approach maintains effectiveness under realistic noise conditions, with fundamental landscape structures persisting despite hardware imperfections.
This resilience enables practical deployment on current NISQ devices.
We identified energy scale as the primary factor affecting cross-problem transferability and developed a straightforward normalization technique that enables effective parameter transfer even between problems with vastly different coefficient magnitudes.
Compared to sophisticated parameter setting strategies like INTERP and FOURIER, our approach reduces computational overhead by orders of magnitude, particularly for deeper circuits, while maintaining comparable solution quality.
The transferability extends across different problem classes, including random Ising models, SK models, and max-cut problems making our approach broadly applicable to diverse optimization challenges.

These findings reshape quantum optimization strategies for near-term devices by prioritizing computational efficiency and transferability over marginal gains in solution quality.
This approach enables addressing larger problems with dramatically reduced classical overhead, a crucial consideration for demonstrating practical quantum utility in the NISQ era.

Several promising research directions emerge from this work.
Creating adaptive algorithms that automatically detect the appropriate energy scale normalization factor between different problem instances would further enhance the approach's versatility.
Indeed, in this paper $|E_{\mathrm{SA}}|/\sqrt{n_{\mathrm{edges}}}$ was used for a heuristic normalization factor, so checking how other criteria are tolerable would facilitate actual use cases.
Establishing rigorous guarantees on the performance gap between linearized and fully optimized parameters for specific problem classes could identify scenarios where linear parameters are provably reliable.
Adapting the approach to handle realistic constrained optimization problems, as recently demonstrated in specific application domains~\cite{Leclerc:2024qra,Priestley:2025vcx}, would extend its applicability to more complex scenarios.
Exploring synergies between parameter transfer and classical optimization techniques like local search or genetic algorithms could create more powerful hybrid approaches that leverage the strengths of both quantum and classical computing.
Indeed, parameter transfer can possibly speed up the recent proposed sampling error mitigation algorithm, NDAR~\cite{arxiv2404},
since it originally involves QAOA parameter optimization at each iteration step:
parameter transfer can eliminate this nested optimization loop.

\appendices

\section{Detail of algorithms for comparison}
\label{sec:interp_fourier}

In the main body of paper, we compare LINXFER (Alg.~\ref{alg:linxfer}) to the established parameter setting methods INTERP and FOURIER.

For reproducibility, we declare the detail of INTERP and FOURIER in Algs.~\ref{alg:interp} and~\ref{alg:fourier}, respectively.
Note that, in the original paper~\cite{Zhou:2018fwi}, the BFGS algorithm is adopted as the classical optimization method in QAOA while we exclusively employ COBYLA in our current paper.
Another important note is that we limit the number of function evaluations to $1000$.
As for the number of Fourier terms in FOURIER, we use $k=2$ for the comparison purpose to LINXFER;
\textit{i.e.} the number of tunable parameters is $4$ in both LINXFER and FOURIER under this choice.
Although the number of parameters is set to be same, the computational complexity of FOURIER is much more demanding than LINXFER, where any instance-dependent optimization is not needed.

\begin{algorithm}
  \caption{INTERP: Parameter Interpolation Strategy~\cite{Zhou:2018fwi}}
  \begin{algorithmic}[1]
    \Require Hamiltonian $H$, number of layers $p$
    \Ensure Optimized QAOA parameters $\gamma$ and $\beta$

    \Function{Interpolate}{$params$}
      \State $p \gets \Call{Len}{params} + 1$
      \State $tmp \gets [0.0] + params + [0.0]$
      \State $ret \gets [\:]$ \Comment{Empty list}
      \For{$i = 1$ to $p$}
        \State $r \gets (i - 1)/p$
        \State Add $r \cdot tmp[i-1] + (1-r) \cdot tmp[i]$ to $ret$
      \EndFor
      \State \Return $ret$
    \EndFunction

    \State $\gamma \gets [0.1]$, $\beta \gets [0.1]$ \Comment{Initial parameters}

    \For{$i = 1$ to $p$}
      \State $\gamma_{\mathrm{init}} \gets \gamma$, $\beta_{\mathrm{init}} \gets \beta$
      \State $\gamma, \beta \gets$ \Call{QAOA}{$H$, $i$}.\Call{optimize}{$\gamma_{\mathrm{init}}$, $\beta_{\mathrm{init}}$} \Comment{Using COBYLA, max $1000$ evals}
      \State $\gamma \gets \Call{Interpolate}{\gamma}$, $\beta \gets \Call{Interpolate}{\beta}$
    \EndFor

    \State \Return $\gamma[0:p], \beta[0:p]$
  \end{algorithmic}
  \label{alg:interp}
\end{algorithm}

\begin{algorithm}
  \caption{FOURIER: Setting w/ Fourier terms~\cite{Zhou:2018fwi}}
  \begin{algorithmic}[1]
    \Require Number of Fourier terms $k$, Hamiltonian $H$, number of layers $p$
    \Ensure Optimized QAOA parameters $\gamma$ and $\beta$

    \Function{Convert}{$u$, $v$}
      \State $\gamma \gets [\:]$, $\beta \gets [\:]$ \Comment{Initialize empty lists}
      \For{$i = 1$ to $p$}
        \State $\gamma_i \gets \sum_{j=1}^{|u|} u_{j-1} \sin\left( (j-0.5)(i-0.5)\pi / p \right)$
        \State $\beta_i \gets \sum_{j=1}^{|v|} v_{j-1} \cos\left( (j-0.5)(i-0.5)\pi / p \right)$
        \State Add $\gamma_i$ to $\gamma$ and $\beta_i$ to $\beta$
      \EndFor
      \State \Return $\gamma, \beta$
    \EndFunction

    \Function{Optimize}{$u_{\mathrm{init}}$, $v_{\mathrm{init}}$}
      \State $f(u, v)$: $\expval{H}$ using $\gamma, \beta \gets$ \Call{Convert}{$u$, $v$}
      \State \Return \Call{Minimize}{$f$, $u_{\mathrm{init}}$, $v_{\mathrm{init}}$} \Comment{Using COBYLA, max $1000$ evals}
    \EndFunction

    \State $u \gets [\:]$, $v \gets [\:]$ \Comment{Initialize empty Fourier coefficient lists}

    \For{$i = 1$ to $k$}
      \State $u_{\mathrm{init}} \gets u + [0.0]$, $v_{\mathrm{init}} \gets v + [0.0]$
      \State $u, v \gets$ \Call{Optimize}{$u_{\mathrm{init}}$, $v_{\mathrm{init}}$}
    \EndFor

    \State \Return $\Call{Convert}{u, v}$
  \end{algorithmic}
  \label{alg:fourier}
\end{algorithm}

\section{Bond dimension dependence of MPS result}
\label{sec:bond_dim_dependence}

The MPS simulator allows us to handle circuit sizes beyond what state vector simulators can deal with in terms of both CPU time and memory requirements,
while introducing its own characteristic errors from the truncated bond dimension $\chi$.

Figure~\ref{fig:bond_dim_dependence_nqubits32_dedges0.6_nsamples8} shows the bond dimension dependence of $\expval{E}/E_{\mathrm{SA}}$ and $E_{\mathrm{best}}/E_{\mathrm{SA}}$ for instances where $n_{\mathrm{qubits}}=32$ and $d_{\mathrm{edges}}=0.6$.
For just putting our focus on the dependence on the bond dimension, we fix the number of layers to $8$ and use a pre-trained parameter set in Eq.~\eqref{eq:optparams_nqubits16_pedge0.6_nlayers8}.
The figure shows a convergence for $\chi \geq 32$, supporting our using the IBM's MPS simulator with $\chi = 32$ in the main body of paper~\footnote{
  Note that the entanglement in the system grows with the number of qubits, the circuit depth, and what kinds of gates are involved.
  Thus it is always difficult to define the ``minimum required bond dimension'' in actual use cases.
  For an entanglement perspective on QAOA, see Ref.~\cite{Dupont:2022qjc}.
}.

A key observation is that, using too rough approximations with $\chi = 2, 4, 8$, the sampled energies $\expval{E}$, $E_{\mathrm{best}}$ are underestimated than those with larger $\chi$s.
Then we suppose the examination of parameter transferability in Sec.~\ref{sec:transfer_scaling} would never be overclaimed.

\begin{figure}[htbp]
  \centering
  \includegraphics[width=\hsize]{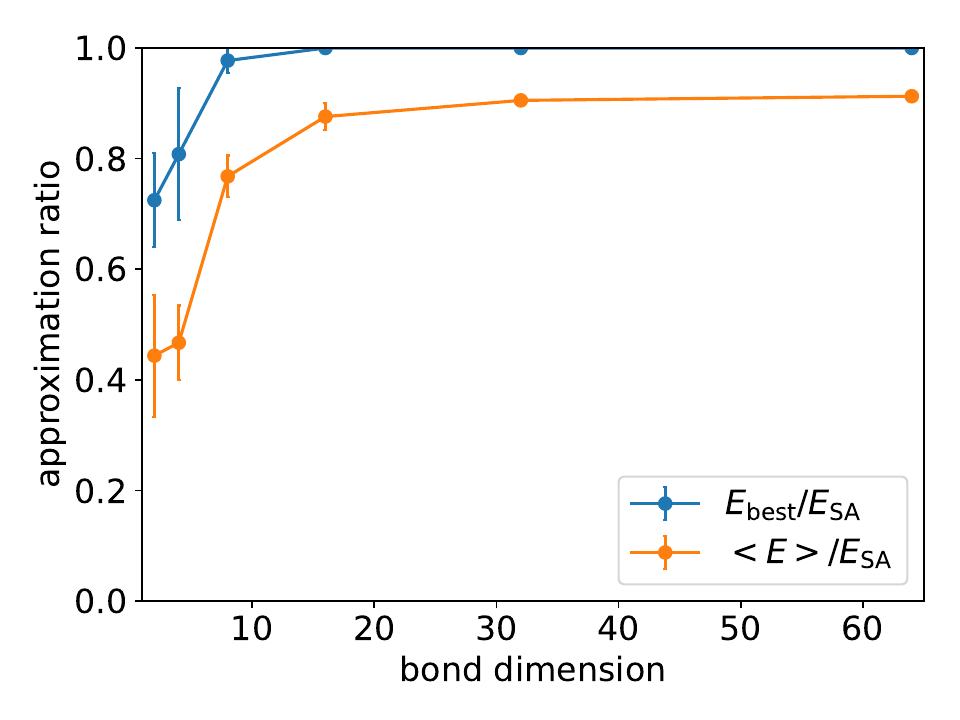}
  \caption{
    Bond dimension dependence of approximation ratios.
    $8$ instances are averaged for each bond dimension.
  }
  \label{fig:bond_dim_dependence_nqubits32_dedges0.6_nsamples8}
\end{figure}

\section*{Acknowledgment}

We acknowledge the use of IBM Quantum Credits for this work.
The views expressed are those of the authors, and do not reflect the official policy or position of IBM or IBM Quantum.

\printbibliography[title=References]

\end{document}